\documentclass[twocolumn,showpacs,preprintnumbers,amsmath,amssymb]{revtex4}
\usepackage{graphicx}
\usepackage{dcolumn}
\usepackage{bm}

\newcommand{\ie}{ i.e.}
\newcommand{\eg}{e.g.}

\newcommand{\half}{\frac{1}{2} }

\def\deriv#1{\frac{\partial}{\partial#1}}
\def\der#1#2{\frac{\partial#1}{\partial#2}}
\def\htau#1{\hat{\tau}^{#1}}

\begin{document}

\title{Understanding spin Hall effects from the motion
in SU(2)$\times$U(1) fields}

\author{Pei-Qing Jin and You-Quan Li}
\affiliation{Zhejiang Institute of Modern Physics and Department of Physics,
Zhejiang University, Hangzhou 310027, P. R. China }

\begin{abstract}
We derive the classical counterpart of a previously obtained quantum
mechanical covariant ``continuitylike'' equation for the spin
density, and present an intuitive picture for elucidating the
non-conservation of the spin current. This reveals the equations of
motion for a particle with spin under the Yang-Mills field (in
certain semiconductors) as well as the Maxwell field, from which the
condition for an infinite spin relaxation time is drawn out
directly. As a concrete example, we discuss the procession of the
spin orientation in spin Hall effect with the so called ReD field,
which undergoes a circle with the frequency dependent on both the
strength of the spin-orbit coupling and the initial velocity. The
anti-commutation of the Pauli matrices is found to be crucial in
simplifying the equations of motion in the view of quantum mechanism
of the same topics.
\end{abstract}


\received{\today}

\maketitle

The spin Hall effect~\cite{Daykonov,Hirsch} is regarded to be
practically useful for the rapidly developing field of
spintronics~\cite{Zutic04}, in which the spin manipulation,
polarization as well as the detection play crucial roles.
Theoretically, the spin Hall effect was predicted for p-type
semiconductors~\cite{ZhangSC} where the up and down spin particles
drift in opposite directions due to an "effective magnetic field"
originating from the Berry phase curvature, and a constant spin Hall
conductivity was speculated for n-type semiconductors~\cite{Niu0403}
with the help of either the Bloch equations or the Kubo formula.
Although the absence of the Hall voltage makes it difficult to
detect the pure spin Hall effect directly, the spin accumulation in
nonmagnetic semiconductors has been observed experimentally by two
groups~\cite{Wunderlich,Awschalom} recently.

The nonconservation~\cite{Niu0407,Li-current} of the spin density in
the presence of spin-orbit couplings brings about some difficulties
relevant to the theoretical analyses~\cite{Li-current,Niu05,Ma},
such as the definition of the spin current. A covariant form for the
continuitylike equation for spin density and current was
given~\cite{Li-current} in the terminology of Yang-Mills gauge
potentials, and it was shown~\cite{Li-Kuboformula} to paly an
essential role in guaranteeing the consistency of a generalized Kubo
formula for the linear response to non-Abelian fields. Since the
work mentioned above are all based on quantum mechanics, one may ask
what is the classical counterpart of the covariant form of the
continuitylike equation for the spin density, and what is the
classical interpretation for the non-conservation of spin current?

For a semiclassical understanding of the spin Hall effects in a
two-dimensional electron system with Rashba coupling, an external
electric field was shown to exert a transverse
force~\cite{Hirsch,Nikolic,Shen} on a moving spin, where the
calculation in Ref.~\cite{Shen} is done on the basis of Ehrenfest
principle and the discussion in Ref.~\cite{Nikolic} is carried out
with the help of an intuitive picture that the center of a wave
packet gets deflection in the transverse direction. The general form
of the forces acting on spin and spin current for a variety of
systems considered in current literatures was given in
Ref.~\cite{Li-current}. The time-dependent Rashba spin-orbit
coupling was ever proposed to create a force acting on opposite
spins in opposite directions~\cite{Malshukov}. This makes it
possible to produce the Yang-Mills field in certain semiconductors
according to the viewpoint of our formulism~\cite{Li-current}. It is
natural to ask whether our formulism has advantage in making sense
of the roles that spin and charge play in spin Hall effect. It is
therefore an emergent task to elucidate the aforementioned issues
via a unified picture.

In this letter, we investigate the motion of a particle with spin in
the Yang-Mills field from a classical point of view. We start with a
revisit to the classical counterpart of the continuitylike equation
for the spin-current density. Such a topic was ever investigated by
various authors~\cite{Sun} but has not been correctly exposed yet.
We derive the same form as ever proposed quantum-mechanically in the
previous paper~\cite{Li-current}, which presents a clear picture of
the non-conservation of the spin current. On the basis of this
result, we obtain the equations of motion for a particle with spin
in the presence of the Yang-Mills field as well as the Maxwell
field. A straightforward condition for an infinite spin relaxation
time is drawn out. Then we discuss the spin Hall effect as a
concrete example and obtain the trajectory of the spin for the equal
strength of Rashba and Dresselhaus couplings. We also discuss the
quantum counterpart of the equations of motion and indicate that
the anti-commutation relation of the Pauli-matrices makes the
problem easy to solve.

\begin{figure}[t]
\includegraphics[width=66mm]{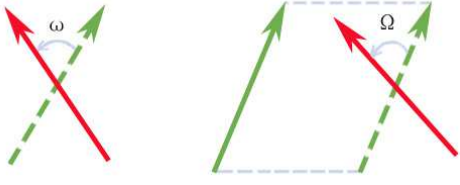}
\caption{\label{fig:procession} (color on line) The left scheme
depicts the time-evolution of $\vec{N}$ at a point $\mathbf{r}$; the
right scheme illustrates the spatial deviation of $\vec{N}$'s by
comparing the fields at two neighborhood points $\mathbf{r}$ and
$\mathbf{r}+\Delta x_i$ for which the parallel translation is
inevitable.}
\end{figure}

We start with considering a moving top (classical analogy of spin)
which rotates at a certain rate. A continuum constituted by such
kind of tops is completely characterized by a local velocity field
$\mathbf{v}(\mathbf{r}, t)$, a local particle-density field
$\rho(\mathbf{r}, t)$ together with a local alignment field
$\vec{N}(\mathbf{r}, t)$
\cite{footnote}.
As illustrated in Fig.~(\ref{fig:procession}),
the time evolution of $\vec{N}(\mathbf{r}, t)$ is determined by
comparing $\vec{N}$ at different times at the same place,
\ie,
$\vec{N}(\mathbf{r}, t+\Delta t)-\vec{N}(\mathbf{r}, t)
 =\vec{\omega}\times \vec{N}(\mathbf{r}, t)\Delta t$, while
its spatial deviation is
determined by comparing
$\vec{N}$ at different places simultaneously,
\ie,
$\vec{N}(\mathbf{r}+\Delta
x_i, t)-\vec{N}(\mathbf{r},t)
 =\vec{\Omega}_i\times \vec{N}(\mathbf{r}, t)\Delta x_i$.
Hereafter, the Latin indices run from $1$ to $3$, and the repeated
indices are summed over. The vector fields $\vec \omega$ and $\vec
\Omega_i$ are natural consequences of $\vec{N}$ being a unit vector,
then we have
\begin{eqnarray}\label{eq:omegafields}
\deriv{t}\vec{N}(\mathbf{r}, t)
  =\vec{\omega}(\mathbf{r}, t)\times\vec{N}(\mathbf{r}, t),
  \nonumber\\
\deriv{x_i}\vec{N}(\mathbf{r}, t)
 =\vec{\Omega}_i(\mathbf{r}, t)\times\vec{N}(\mathbf{r}, t).
\end{eqnarray}
By making use of these two relations together with the density
conservation $\displaystyle \der{\rho}{t}  + \der{j^{~}_i}{x_i}=0$, we
obtain a continuitylike equation
\begin{equation}\label{eq:continuty-like}
\bigl(\deriv{t}-\vec{\omega}\times\bigr)\vec{\sigma}
 +\bigl(\deriv{x_i}-\vec{\Omega}_i\times\bigr)\vec{J}_i=0,
\end{equation}
as long as the nature definitions of spin density
$\vec{\sigma}=\rho\,\vec{N}$ and spin-current density
$\vec{J}_i=\rho\,v_i\,\vec{N}$ are employed. Comparing with the
quantum mechanical results ~\cite{Li-current}, one can recognize
that $\vec{\omega}$ and $\vec{\Omega}_i$ correspond to the
Yang-Mills gauge potentials $\eta\vec{\mathcal A}_0$ and
$-\eta\vec{\mathcal A}_i$, respectively, which have been
shown~\cite{Li-current}  to describe the spin-orbit coupling such as
Rashba~\cite{Rashba}, Dresselhaus~\cite{Dresselhaus} coupling and
etc. As we are aware, a clear and direct physical meaning of the
non-conservation of spin current has not been exposed before.

The above results are suggestive for us to elucidate the equations
of motion for a charged vector $\vec{n}$ of constant magnitude
(classical analogy of a charged particle with spin) moving in the
space undergoing both the Maxwell and Yang-Mill fields
\begin{eqnarray}\label{eq:equationofmotion}
\frac{d\vec{n}(t)}{dt} = \eta(\vec{\mathcal A}_0
 -v^{}_i\vec{\mathcal A}_i)\times \vec{n}(t), \hspace{37mm}
 \nonumber\\
m\frac{d v^{}_i}{dt} = \vec{\mathcal E}_i\cdot\vec{n}(t)+e E_i
  +\epsilon^{}_{ijk}v^{}_j(\,\vec{\mathcal B}_k\cdot\vec{n}(t)+e B_k).
   \hspace{8mm}
\end{eqnarray}
The first equation for the spin orientation is obtained
from Eq.(\ref{eq:omegafields}) by adopting $\vec{N}(\mathbf{r}, t) =
 \vec{n}(t)\delta(\mathbf{r}-\tilde{\mathbf r}(t))$ with
$\tilde{\mathbf r}(t)$ being the trajectory of a single particle.
The second equation is due to the fact that the translational motion
is governed by both the Lorentz force caused by the Maxwell fields
$E_i$ and $B_i$ and the force by the Yang-Mills fields
$\vec{\mathcal E}_i$ and $\vec{\mathcal B}_i$ which are
matrix-valued vectors~\cite{Li-current}. Hereafter, $\vec{n}(t)$ is
written as $\vec{n}$ for simplicity. To avoid ambiguity, we need to
set up a picture with two distinct spaces. One is the SU(2) Lie
algebra space (we call it spin space hereafter) in which the
Yang-Mills gauge potentials and fields are defined, the other is the
conventional spatial space. The ``coordinate bases'' of the former
are $\{\htau{1}, \htau{2}, \htau{3} \}$ with $2\hat{\tau}$ referring
to the Pauli matrices, \eg, $\mathbb{B}_i=\mathcal{B}_i^1 \htau{1}
   +\mathcal{B}_i^2 \htau{2}
     +\mathcal{B}_i^3 \htau{3},$
while those of the later are $\{\mathbf{e}_x, \mathbf{e}_y,
\mathbf{e}_z\}$, \eg, $\mathbf{v}=v_x
\mathbf{e}_x+v_y\mathbf{e}_y+v_z \mathbf{e}_z$. In the language of
the gauge potentials, the Yang-Mills ``electric'' and ``magnetic''
fields can be expressed as
\begin{eqnarray}
\vec{\mathcal E}_i &=& -\partial_0 \vec{\mathcal A}_i
 -\partial_i \vec{\mathcal A}_0 +\eta\vec{\mathcal A}_0\times\vec{\mathcal A}_i,
  \nonumber\\
\vec{\mathcal B}_i &=& \epsilon_{ijk}\partial_j \vec{\mathcal A}_k
         +\frac{\eta}{2}\epsilon_{ijk}\vec{\mathcal A}_j\times\vec{\mathcal A}_k.
\end{eqnarray}
It is worthwhile to note that the non-Abelian fields can be
non-vanishing even when the gauge potentials are constant.

In the view of the analytical mechanics, the equations of
motion~(\ref{eq:equationofmotion}) can be formulated in the
Hamiltonian formulism with $
H=\frac{1}{2m}(p^{}_i-eA_i-\eta\vec{\mathcal A}_i\cdot\vec n)^2
 +eA_0+\eta\vec{\mathcal A}_0\cdot\vec n.$
As a standard procedure, one needs to choose two sets of canonical
coordinates and their conjugations (canonical momentums), namely,
$(\phi, n_z)$ with $\phi=\tan^{-1}(-n_x/n_y)$ and $(r_i, v_i)$. One
pair of the canonical equations can be easily derived
\begin{eqnarray*}
\frac{d \phi}{dt} & = & \frac{\partial H}{\partial n_z}
     =\frac{\eta}{n_\bot^2}
     \bigl( (\mathcal A_0^z \!-v_i\mathcal A_i^z)n^2
           - n_z(\vec{\mathcal A}_0-v_i\vec{\mathcal A}_i)\cdot\vec{n}
     \bigr),
    \nonumber \\
\frac{d n_z}{dt} &=& -\frac{\partial H}{\partial \phi}
 =\bigl( \eta (\vec{\mathcal A}_0-v_i\vec{\mathcal A}_i)\times\vec n
  \bigr)_z,
\end{eqnarray*}
with $n_\bot^2=n^2 - n_z^2$, which is exactly the first equation of
Eq.~(\ref{eq:equationofmotion}) after some algebra. The other pair
of the canonical equation respect to $(r_i, v_i)$ can be shown to be
the second one of Eq.~(\ref{eq:equationofmotion}).

Some immediate consequences can be obtained from
Eq.~(\ref{eq:equationofmotion}). ({\bf i}) The first equation for
the time rate of $\vec{n}$ clearly manifests that the vector $\vec
n$ does not process when it is parallel to $\vec{\mathcal A}_0 -
v_i\vec{\mathcal A}_i $, or more specially, $\vec{\mathcal A}_0 -
v_i\vec{\mathcal A}_i=0$. In such cases, the spin orientation keeps
unchanged, which results in an infinite spin relaxation time. A
typical example is that the infinite spin relaxation time occurs in
the $\pm[1,\pm1,0]$ direction when $\vec{\mathcal A}_0=0$ and
$\alpha=\pm \beta$ ($\alpha$ and $\beta$ refer to Rashba and
Dresselhaus coupling strength, respectively, in the quantum case),
one of which was discussed in Ref.~\cite{SCZhang06}. This is
analogous to the case in the classical electrodynamics where an
electron moving in the uniform orthogonal electromagnetic fields
with velocity $\mathbf{v} = \mathbf{E}\times\mathbf{B}/B^2 $ does
not feel the Lorentz force. ({\bf ii}) In virtue of the coupling
between $\vec n$ and the Yang-Mills field in the second equation of
(\ref{eq:equationofmotion}), the time rate of $\vec n$ leads to the
time-dependent effective fields even when the Yang-Mills fields are
time-independent, which is quite different from the motion in the
Maxwell field.

\begin{figure}[h]
\includegraphics[width=66mm]{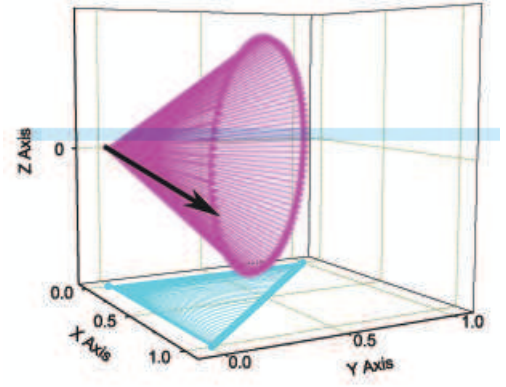}\\
\caption{(color on line) The rotation of $\vec n$ is depicted with the initial
conditions $\vec n_{0}=(1,0,0)$, where the x-y projection
is also shown to elucidate this motion.}\label{fig:rotation}
\end{figure}

For many real physical systems, the gauge potentials $\vec{\mathcal
A}_i$ are constant. If we focus on the contribution of the
Yang-Mills field and set the Maxwell one to be zero, the second
equation of (\ref{eq:equationofmotion}) reduces to
\begin{equation}
m\frac{dv_i}{dt} = -\eta\vec{\mathcal A_i}\cdot(\vec{\mathcal A}_0 -
v_j\vec{\mathcal A}_j)\times \vec{n},
\end{equation}
which gives rise to a relation between $\vec n$ and $v_i$
\begin{equation}\label{eq:relation between n and v}
v_i(t)=-\frac{1}{m}\vec{\mathcal A}_i\cdot\vec n(t)+C_i,
\end{equation}
where $C_i=v_{0 i}+\frac{1}{m}\vec{\mathcal A}_i\cdot\vec n_0$ are
determined by the initial values $v_i(0)=v_{0 i}$ and $\vec
n(0)=(n_{0 x},n_{0 y},n_{0 z})$. Consequently, one only needs to
solve one equation,
\begin{equation}\label{eq:motion for n}
\frac{d\vec n}{dt}=\eta(\vec{\mathcal A}_0-C_i\vec{\mathcal
A}_i)\times\vec n
 +\frac{\eta}{m}(\vec{\mathcal A}_i\times\vec n)
  (\vec{\mathcal A}_i\cdot\vec n).
\end{equation}
As a concrete example, we consider a two-dimensional system with
uniform Yang-Mill gauge potentials~\cite{Li-current}
\begin{eqnarray}\label{eq:potentials}
 \vec{\mathcal A}_x \!=\!\frac{2m}{\eta^2}(\beta,\,\alpha,\,0), \hspace{3mm}
 \vec{\mathcal A}_y \!=\!-\frac{2m}{\eta^2}(\alpha,\, \beta,\,0), \hspace{3mm}
 \vec{\mathcal A}_0 \!=\!0,   \hspace{5mm}
\end{eqnarray}
which correspond to Rashba and Dresselhaus couplings in certain
semiconductors. Accordingly, the equations of motion are explicitly
written as
\begin{eqnarray}
\frac{d n_x}{dt} = -\frac{2m}{\eta} n_z
   \Bigl[\alpha C_1-\beta C_2
   -\frac{4\alpha\beta n_x  + 2(\alpha^2\!+\!\beta^2) n_y }{\eta^2}
   \Bigr],
  \nonumber\\
\frac{d n_y}{dt} = \frac{2m}{\eta} n_z
   \Bigl[\beta C_1 - \alpha C_2
    -\frac{4\alpha\beta n_y + 2(\alpha^2\!+\!\beta^2)n_x }{\eta^2}
    \Bigr], \hspace{3mm}
  \nonumber\\
\frac{d n_z}{dt} = \frac{2m}{\eta} n_x
   \Bigl[\alpha C_1 - \beta C_2
      - \frac{4\alpha\beta n_x}{\eta^2}
      \Bigr] \hspace{28mm}
       \nonumber\\
  - \frac{2m}{\eta} n_y
    \Bigl[C_1\beta - C_2\alpha
   -\frac{4\alpha\beta n_y}{\eta^2}
   \Bigr]. \hspace{24mm}
\end{eqnarray}
For $\alpha=\beta$ which is called ReD field in
Ref.~\cite{SCZhang06}, we can solve these equations analytically:
\begin{eqnarray}\label{eq:n-0th order}
 n_x \!\!&=&\!\! -a\sin(\omega t+\varphi)+\frac{1}{2}(n_{0 x}+n_{0y}),
 \nonumber \\
 n_y \!&=& a\sin(\omega t+\varphi)+\frac{1}{2}(n_{0x}+n_{0y}),
 \nonumber \\
 n_z &=& \sqrt{2}a\cos(\omega t+\varphi),
\end{eqnarray}
where $a=\sqrt{n_{0z}^2/2 + (n_{0x}-n_{0y})^2/4}$ and $\tan
\varphi=(n_{0y}-n_{0x})/(\sqrt{2}\,n_{0z})$ are determined by the
initial conditions. It is clear that the tip of $\vec n$ experiences
a cyclotron rotation with frequency
$\omega=\frac{2\sqrt{2}m\alpha}{\eta}(v_{0x}-v_{0y})$, which is
shown in Fig.~(\ref{fig:rotation}) for the initial condition $\vec
n_{0}=(1,0,0)$.
The instantaneous velocity solved from Eq.~(\ref{eq:relation between n and v})
is just its initial value $v_x=v_{0x}$ and $v_y=v_{0y}$,
\ie, the electron undergoes a motion with uniform velocity.
This is due to the time-dependent parts of $n_x$ and $n_y$
only differ from each other by a minus sign. Specially, when
$v_{x}=v_{y}$, the spin vector $\vec n$ does not process since
$\omega=0$, which recovers the result in Ref~\cite{SCZhang06}.

When an external electric field $\vec E=(E_x,E_y)$ is applied, which
mimics to the usual spin Hall effect in current literature, we
obtain $v_i(t)=-\frac{1}{m}\vec{\mathcal A}_i\cdot\vec
n(t)+C_i+\frac{e}{m}E_i t$. The equation of motion for $\vec n$ is
almost the same as Eq.~(\ref{eq:motion for n}) if substituting $C_i$
by $\tilde C_i(t)=C_i+\frac{e}{m}E_i t$. For the electric field is
sufficiently weak, the perturbation theory is applicable and we can
expand the spin unit vector $\vec n$ in the power of the electric
field,
\begin{equation}
\vec n=\vec n^{(0)}+\vec n^{(1)}+\cdots.
\end{equation}
The $0$th-order results take the form of Eq.~(\ref{eq:n-0th order}),
while the linear corrections are
\begin{eqnarray}
 n_x^{(1)} \!\!&=&\!\! -\lambda(E)~ t^2\cos(\omega t+\varphi),
 \nonumber \\
 n_y^{(1)} \!&=& \lambda(E)~ t^2\cos(\omega t+\varphi),
 \nonumber \\
 n_z^{(1)} &=& -\sqrt{2} \lambda(E)~ t^2\sin(\omega t+\varphi),
 \end{eqnarray}
where $\lambda(E)=\alpha e a (E_x-E_y)\sqrt{2}/\eta $. The form of
the entire motion for $\vec n$ is much similar to the case without
external electric fields, except for a time-dependent amplitude
$\tilde a (t)=[a^2+\lambda^2 t^4]^{1/2}$ and a phase $\tilde
\varphi(t)=\varphi+\tan^{-1}(\lambda t^2/a)$. Even though the
rotation of $\vec n$ is a bit of intricacy, its translational motion
only involves one more term linearly dependent on time,
\begin{eqnarray}
v_x &=& v_{0x}+\frac{e}{m}E_x t,
   \nonumber \\
v_y &=& v_{0y}+\frac{e}{m}E_y t.
\end{eqnarray}

Furthermore, we turn to the quantum picture of the motion of a
single particle with spin in the background of the Yang-Mills field
with the Hamiltonian $ \hat{H} = \frac{1}{2m}
  \bigl(\hat{p}_i-eA_i- \eta\vec{\mathcal A}_i\cdot\vec\tau\bigr)^2
   + e A_0 + \eta \vec{\mathcal A}_0\cdot\vec\tau
  $~\cite{Li-current}.
The time rate of its dynamical momentum $\hat\pi_i \equiv \hat p_i-
eA_i -\eta\vec{\mathcal A}_i\cdot\vec\tau$ is given by the
Heisenberg equation of motion $d\hat\pi_i/dt\!\!=\!\!
\frac{1}{i\hbar}[\hat\pi_i,H]
 = eE_i+\eta\vec{\mathcal E_i}\cdot{\vec\tau}
  +\half\epsilon^{}_{ijk}
  \{v_j,eB_k+\eta\vec{\mathcal B_k}\cdot\vec\tau \}$,
where the curve parentheses denote the anti-commutator.
Similarly, the equation of motion for the spin operator
reads
$ d \vec\tau / dt =\frac{1}{i\hbar}[\vec\tau,H]
 =\eta\vec{\mathcal A}_0\times\vec\tau
  -\frac{\eta}{2}\{v_j, \vec{\mathcal A}_j\times\vec\tau \}$.
Obviously, these equations are the quantum counterpart of
Eq.~(\ref{eq:equationofmotion}) with a $\eta\vec\tau$ to $\vec n$
correspondence. In the previous classical case, it is difficult to
solve the equations of motion since they are coupled to each other.
However, things become easier in the quantum case. Substituting $
v^{}_j=\hat\pi_j/m=\frac{1}{m}(p_j-eA_j-\eta\vec{\mathcal
A}_j\cdot\vec\tau) $ into the equation of motion for $\vec\tau$, we
obtain $ d\vec\tau / dt = \eta\vec{\mathcal A}_0\times\vec\tau
-\frac{\eta}{2}\{p_j-eA_j, \vec{\mathcal A}_j\times\vec\tau \}$ in
which the anti-commutation relation
$\{\tau^a,\tau^b\}=\frac{1}{2}\delta^{ab}$ has been used. The time
evolution of $p_j$ obeys $dp_j/dt=[p_j,H]/i\hbar$. The equations of
motion which are coupled to each other in the classical mechanics
become decoupled in the quantum case where the anti-commutation
relations of $\vec\tau$ play a key role. For example, in the
conventional spin Hall system, the external electric field is
applied to drive the spin current and $dp_i/dt = e E_i$. The
equation of motion for $\vec\tau$ reduces to the one ever considered
in Ref.~\cite{Niu0403}.

In the above, we investigated the motion of a particle with spin in
the Yang-Mills field from a classical point of view which differs
from that of Ref~\cite{Sun}, and presented a much more clear picture
for the non-conservation of spin current, which has not been
correctly exposed before. We revisited to the classical counterpart
of continuitylike equations for spin-current density, which has the
same form as we ever proposed quantum-mechanically in a previous
paper~\cite{Li-current}. With those results, we elucidated the
equations of motion for a particle with spin in the presence of
Yang-Mills fields as well as the Maxwell fields. Although various
authors~\cite{Niu0403} have considered the equation of motion, the
complete form (\ref{eq:equationofmotion}) has not been recognized
yet. It is worthwhile to note that the appearance of the coupling
between the spin orientation and the Yang-Mills fields brings about
the time-dependent effective fields even when the Yang-Mills fields
are time-independent. In other words, the motion of the particle is
greatly affected by the procession of its spin. From one of
equations of motion, we can easily obtain the condition when the
spin relaxation time is infinite and a special case of our
conclusion recovers the previous result discussed by other authors.
We have taken the spin Hall system as a concrete case and found that
the tip of the spin vector undergoes a cyclotron motion with the
frequency determined by the strength of the spin-orbit coupling and
the initial velocity of the particle. Furthermore, we reviewed the
quantum counterpart of the equations of motion and indicated that
the anti-commutation relation of the Pauli matrices makes the
equations of motion decoupled, which is thus easy to solve.

The work was supported by NSFC grant No.10225419.


\begin{references}
\bibitem{footnote}
A complete  quantum mechanical description is given by four local quantities
$\psi^*_\uparrow \psi_\uparrow$, $\psi^*_\uparrow \psi_\downarrow$,
$\psi^*_\downarrow \psi_\uparrow$, and $\psi^*_\downarrow \psi_\downarrow$,
of which the proper compositions define one probability
density and three components of  spin densities,
whose classical counterparts are $\rho$ and $\vec{n}$.
\end{references}
\end{document}